\documentclass[10pt, a4paper,twocolumn]{article}

\usepackage{graphicx}% Include figure files
\usepackage{commath}
\usepackage{amsthm}
\usepackage{hyperref}
\usepackage{color}
\usepackage[table]{xcolor}% http://ctan.org/pkg/xcolor
\usepackage{multirow}
\usepackage{booktabs}
\usepackage{subfigure}
\usepackage{float}
\usepackage{authblk}

\usepackage[english]{babel}
\usepackage{amssymb,amsmath,amsthm, amsfonts}

\RequirePackage[left=2cm,%
                right=2cm,%
                top=2.25cm,%
                bottom=2.25cm,%
                headheight=12pt,%
                letterpaper]{geometry}%

\title{Chaotical PRNG based on composition of logistic and tent maps using deep-zoom}

\author[1]{Jo\~ao Pedro do Valle Alvarenga}
\author[1]{Jeaneth Machicao}
\author[1,*]{Odemir M. Bruno}
\affil[1]{\small{S\~{a}o Carlos Institute of Physics, University of S\~{a}o Paulo (USP), PO Box 369, 13560-970, S\~{a}o Carlos, SP, Brazil. \protect\\Scientific Computing Group - http://scg.ifsc.usp.br}}
\affil[*]{\small{bruno@ifsc.usp.br}}

\date{}
\begin{document}
\maketitle
%\noindent{$^1$Scientific Computing Group. S\~{a}o Carlos Institute of Physics, University of S\~{a}o Paulo, S\~{a}o Carlos - SP, PO Box 369, 13560-970, Brazil.}

\begin{abstract}
We proposed the deep zoom analysis of the composition of the logistic map and the tent map, which are well-known discrete unimodal chaotic maps. The deep zoom technique transforms each point of a given chaotic orbit by removing its first $k$-digits after the fractional part. We found that the pseudo-random qualities of the composition map as a pseudo-random number generator (PRNG) improves as the $k$ parameter increases. This was proven by the fact that it successfully passed the randomness tests and even outperformed the $k$-logistic map and $k$-tent map PRNG. These dynamical properties show that using the deep-zoom on the composition of chaotic maps, at least on these two known maps, is suitable for better randomization for PRNG purposes as well as for cryptographic systems.
\end{abstract}

\section{Introduction}
%where chaos is the mathematical machinery 
%, where there is a need to obtain sources of pseudo-randomness in chaotic systems with high positive Lyapunov exponent
%motivation of chaos theory

Pseudo-random number generators (PRNG) are of great importance in various fields as well as applications, especially in simulation, gambling, language programming, cryptography and other areas. In this sense, many efforts have been focused on the use of chaos to build PRNGs by using them as random-like time series, because of the sensitivity to the initial conditions, the fact that complex behavior can emerge even from simple equations, and their unpredictability, i.e. the difficulty to predict over a long period of time, which represent the main mathematical machinery for developing chaos-based PRNGs.

%Chaos theory is a branch of mathematics that is still in its infancy. Since its introduction no more than 55 years ago, after the study of a set of equations proposed by Edward Lorenz, the term ``butterfly effect'' appeared. 

%literature review of randomization: e.g composition of maps, discretization, skiping, deep-zoom
The main approach to constructing chaos-based PRNGs is to build chaotic maps with initial conditions (``seeds'') to start the system and obtain outputs that have uniform distribution, uncorrelated series, ease of implementation and other criteria \cite{RobustPRNG, kietzmann2021guideline}. Certainly, those criteria are a good skeleton of a chaos-based PRNG, however there is still need to enhance the randomness properties of these PRNGs.

In literature there are various works that focused on the randomness enhancement, however the randomization is not an easy task. 
%for instances,  contrary to what one might expect, the approach using the shuffle of two chaotic signals is not effective because xxxxxxxxxxxxxxxxxxxxx. 
Some attempts have combined chaotic maps \cite{alawida2019enhancing} that alternate according to a rule, and still other approaches use randomization techniques to enhance chaos, such as the symbolic dynamics method, for instances, \textit{discretization} and \textit{skipping} \cite{Gonzalez2005}. Other approaches use the composition of maps \cite{Lambi2015, 6470984, cripto-composed-map, Sayed2017}, and the combinations of chaotic maps \cite{alawida2019enhancing, ali2020novel, pisarchik2010chaotic}, mainly for cryptographic purposes. As for the strategy of composing maps, it has been used in cryptography mainly for \cite{cripto-composed-map}, other attempts showed ways to break cryptographic systems based on the composition of chaotic maps \cite{break-composed-maps}. 
%A recent randomization method based on the transformations of the orbits, called as \textit{deep-zoom} \cite{Machicao17} shown interesting results because it was theoretically demonstrated the high increase of the Lyapunov exponent, which is a measure of chaos, as the $k$ parameter is increased \cite{Machicao2021}. 

%deep zoom 
Recently, Machicao \& Bruno \cite{Machicao17} proposed the deep-zoom approach as a randomization method to improve the random-like properties of chaotic maps. Besides that, it has been  theoretically demonstrated the high increase of the Lyapunov exponent, a measure of chaos, as the $k$ parameter is increased \cite{Machicao2021}. This method receives as input a chaotic orbit $\mathcal{O}\{x_0, x_1,\ldots, x_t\}$ given by a chaotic map, with its parameters and initial conditions, so that a new orbit $\mathcal{O}^{k}\{x_0^k, x_1^k,\ldots, x_t^k\}$ is returned as output. This new $\mathcal{O}^k$-orbit consists of transforming each $x_i$ value by removing its $k$-digits to the right of the separator. The authors have demonstrated that increasing the $k$ parameter improves the random quality of the proposed PRNGs based on well-known chaotic maps such as the logistic and the tent map.

%logistic map and tent map
Therefore, in our study, we want to investigate the dynamic behavior of the composition of two well-known chaotic maps. First, the logistic map given by Equation \eqref{eq:Logistic}, which has been proposed as dynamic population model where $x_t$ represents the population in the $t$-th generation. This map shows a surprising range of behaviors as the growth rate parameter $\mu$ is varied. 
\begin{equation}
 f(x_t) = x_{t+1} = \mu x_t (1 - x_t)\,,
\label{eq:Logistic}
\end{equation} 
\noindent whose phase space is defined in $f:[0, 1] \rightarrow [0, 1]$, and $\mu \in [0, 4]$ is the control parameter. In similar manner, another well-known unimodal chaotic map is the tent map given by the Equation~\ref{eq:Tent},
\begin{equation}
g(x_t) = x_{t+1} = 
\begin{cases}
\Gamma x_t,& \text{if } x_t < \frac{1}{2}\\
\Gamma (1-x_t),& \text{if } x_t \geq \frac{1}{2}\,,
\end{cases} 
\label{eq:Tent}
\end{equation}
\noindent where $\Gamma\in [0,2]$ is the control parameter.

%proposta
In this paper, we propose to study the dynamic of the composition of two chaotic maps in a deep-zoom approach to improve their pseudo-randomness properties for PRNG. To this end, we first investigated two ways of composing the logistic map and the tent map, and vice versa, using visualization tools such as bifurcation diagrams, space-time diagrams, Lyapunov exponents, and randomness test such as TestU01 and DIEHARD. In particular, within the chaotic regions, of our interest, we explored the properties of pseudo-randomness using the deep-zoom approach. Thus, we focused on the $k$-orbit sequences to generate sequences of numbers by restricting the parameters, e.g., by fixing $\mu$ to study $\Gamma$, and vice versa.

%PROPOSTA DE PRNG e comparacao

This paper is organized as follows: In section \ref{sec2} we introduce the notations of the composition of logistic and tent maps and vice versa. In section \ref{sec3}, we detail the dynamic analysis of the composition of maps. In section \ref{sec4}, we also analyze the dynamic properties of the composed map using the deep zoom approach. In Section \ref{sec5}, we study the pseudo-randomness of a proposed PRNG based on the composition map, and present the results of the analysis performed using DIEHARD and the TestU01 suite. Finally, in sections \ref{sec6} and \ref{sec7}, discussions and conclusions are presented.
 
%---------------------------------------------- 
\section{Composition of maps FoG and GoF}
\label{sec2}
\subsection{Notations}

Let $f \circ g$ be the composition of the logistic map with the tent map defined by Equation~\eqref{eq:fog}. Similarly, let $g \circ f$ be composition of the tent map with the logistic map given by Equation~\eqref{eq:gof}. 

\begin{equation}
 (f\circ g) (x_{t+1})= 
 \begin{cases}
 \mu \Gamma x_t(1-\Gamma x_t), & \text{if } x_t < \frac{1}{2}\\
 \mu \Gamma (1-x_t)[1-\Gamma(1-x_t)],& \text{if } x_t \geq \frac{1}{2}\,.
 \end{cases}
\label{eq:fog}
\end{equation}

\begin{equation}
 (g\circ f) (x_{t+1})= 
 \begin{cases}
 \Gamma \mu x_t (1 - x_t), & \text{if } x_t < \frac{1}{2}\\
 \Gamma (1- [\mu x_t - (1 - x_t)],& \text{if } x_t \geq \frac{1}{2}\,.
 \end{cases} 
\label{eq:gof}
\end{equation}

From now on, we simply use $FoG$ and $GoF$ to describe both maps respectively. Moreover, these composed maps keep their control parameters, i.e., $\mu \in [0, 4]$ and $\Gamma \in [0,2]$, and that their space-phase is set in $[0, 1] \rightarrow [0, 1]$.

\subsection{Dynamical analysis of FoG}
 
In this section, we have analyzed the general dynamical properties of the orbits generated by the composition map $FoG$ using various plots, including Lyapunov exponents, bifurcation diagrams, frequency distributions and cobweb plots.
 
\subsubsection{Lyapunov exponent stability}
\label{sec:lyapunov}
Regarding the Lyapunov exponent, it can be assessed analytically by using Equation~\eqref{eq:normalLE}
\begin{equation}
\label{eq:normalLE}
\lambda(x_0) = \lim_{T\to\infty}\frac{1}{T}\sum_{t=0}^{T-1} \ln\abs{h(x_t)^\prime}\,,
\end{equation}

\noindent where $h(x_t)^\prime$ is the \textbf{differentiation} of the given function to be analyzed. Thus, to calculate the analytic Lyapunov exponent of the $FoG$ map, we obtained its analytic derivation given by Equation~\eqref{eq:fog_deriv}.

\begin{equation}
 (FoG)' = 
 \begin{cases}
 \mu \Gamma (1-2\Gamma x_t), & \text{if } x_t < \frac{1}{2}\\
 \mu \Gamma (2\Gamma - 2\Gamma x_t -1),& \text{if } x_t \geq \frac{1}{2}\,,
 \end{cases}
\label{eq:fog_deriv}
\end{equation}

In similar manner the analytic Lyapunov exponent for the $GoF$ is given by Equation~\eqref{eq:gof_deriv} 

\begin{equation}
 (GoF)' = 
 \begin{cases}
 \Gamma \mu (1-2x_t), & \text{if } x_t < \frac{1}{2}\\
 -\Gamma (\mu + 1),& \text{if } x_t \geq \frac{1}{2}\,,
 \end{cases} 
\label{eq:gof_deriv}
\end{equation} 

\begin{figure*}[!htbp]
 \centering
\includegraphics[width=\textwidth]{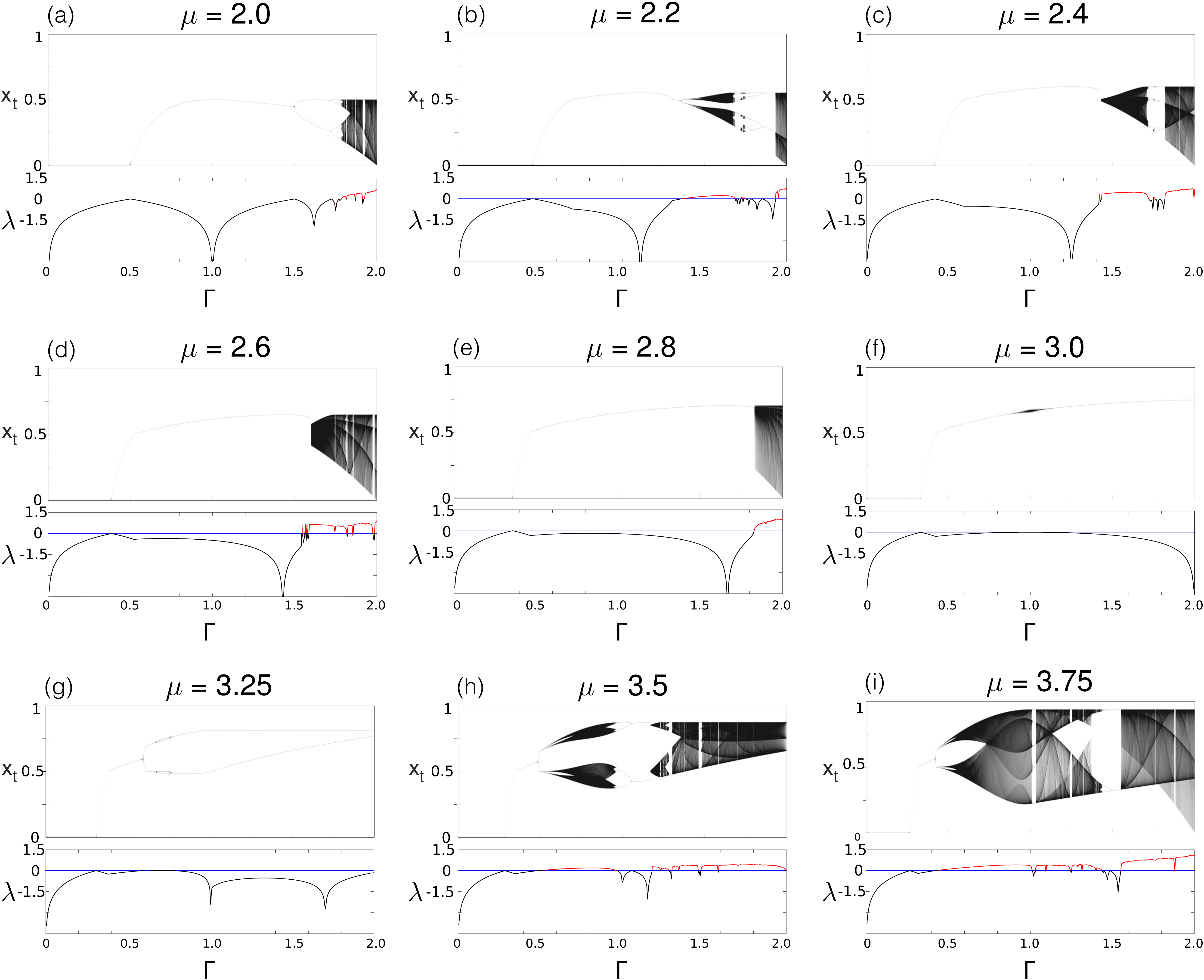}
\caption{Bifurcation diagram (top) and Lyapunov exponent plot (bottom) of $FoG$ map with parameters (a-i) with $\mu \in \{2.00, 2.20, 2.40, 2.60, 2.80, 3.00. 3.25, 3.5, 3.75\}$. The horizontal axis represents the parameter $\Gamma \in [0,2]$. The vertical axis of the bifurcation diagram, $x_t \in [0,1]$, shows the long-term evolution for the same initial condition using the corresponding parameters $\mu$ and $\Gamma$. The system was iterated for $10^5$ iterations and  the first 200 (transient time) were discarded. The lower vertical axis shows the corresponding Lyapunov exponent using the Equation~\eqref{eq:normalLE}. The negative values ($\lambda <0$) are shown in black, while positive Lyapunov values ($\lambda\geq0$) are shown in red. The blue line indicates ($\lambda=0$).}
\label{fig:bif_lya_fog}
\end{figure*}

In Fig.~\ref{fig:bif_lya_fog} (bottom), the Lyapunov exponent of $FoG$ (Equation~\eqref{eq:normalLE}) is plotted using various parameters $\mu\in\{2.00, 2.20, 2.40, 2.60, 2.80, 3.00, 3.25, 3.5, 3.75\}$. The curves are shown so that the negative values ($\lambda <0$) are in black and the positive Lyapunov values ($\lambda\geq0$) are in red. At the top of the Fig.~\ref{fig:bif_lya_fog} shows the bifurcation diagram corresponding to the Lyapunov exponent.

%\textcolor{red}{Detalhar mais elementos da figura xxx xxx xxx xxx xxx xxx xxx xxx xxx}

If we vary the control parameter $\mu$, we can see two regions: two chaotic regions and two stable regions. The chaotic region lies in the interval $\mu \in [1.79, 2.92]$ and $\mu \in [3.25,4.0]$ and the stable one, $\mu\in[0.0, 1.79]$ and $\mu\in[2.92, 3.25]$. These regions are discussed in more detail below.

\subsubsection{Bifurcation Diagram}
\label{sec:biff}
From the point of view of Lyapunov stability and bifurcation of dynamical systems, we aim to understand their stable, periodic and chaotic behavior when their parameters are changed. Thus, by restricting the parameters, for example, by fixing the $\mu$ to study $\Gamma$, and vice versa. 
 
Studying the dynamics of bifurcation diagrams is essential to determine the choice of control parameters for use as PRNGs. We then classify the dynamics of each interval in the bifurcation diagram to facilitate this choice.

\subsubsection{Analysis of region \texorpdfstring{$\mu \in [0, 1.79]$}{Lg}} 
\label{sec:first-interval}

For small values of $\mu$, the bifurcation diagram and the Lyapunov exponent show extremely stable dynamics, i.e., their orbits do not diverge and the Lyapunov exponent consequently has values below zero.

Moreover, this region has extremely concentrated orbits in a certain region of the space-phase, which allows us to claim that their distribution is non-uniform.

Together with the factors: low value of Lyapunov exponent and non-uniformity distribution of orbits in space-phase, this $\mu$ regions is discarded for PRNG.

\subsubsection{Analysis of region \texorpdfstring{$\mu \in (1.79, 2.92]$}{Lg}}

In this range of $\mu$, the FoG map exhibits chaotic behavior confined to high values of the control parameter $\Gamma$. Thus, the Lyapunov exponent is positive only for values of $\Gamma$ larger than $1.5$, as can be seen in Figs.~\ref{fig:bif_lya_fog} (a-e).

The bifurcation diagrams of Figs.~\ref{fig:bif_lya_fog} (a-e) show a distribution of orbits restricted to certain regions within the interval $[0, 1]$. It is extremely complicated to determine precisely the interval that the orbits occupy. Therefore, this region is not ideal for use as a PRNG, because the orbits occupy an imprecise bounding region and the Lyapunov exponent is positive for a small range of the control parameter $\Gamma$.

\subsubsection{Analysis of region \texorpdfstring{$\mu \in (2.92, 3.25]$}{Lg}}

Like the first analyzed interval of $\mu$, this interval has stable orbits (Sec.~\ref{sec:first-interval}), i. e., it has a negative Lyapunov exponent, like the  Figs.~\ref{fig:bif_lya_fog} (f-g). Moreover, unlike the first interval analyzed, the orbits are on two different values, as observed in Fig.~\ref{fig:bif_lya_fog} (g).

Given the negative Lyapunov exponent and the distribution of the orbits, this $\mu$ range is discarded for use as a PRNG. 

\subsubsection{Analysis of region \texorpdfstring{$\mu \in (3.25, 4.0]$}{Lg}}

The last analyzed region has some interesting features that highlight the previously analyzed regions. As the Figs.~\ref{fig:bif_lya_fog}(h-i) show, the Lyapunov exponent is larger than then zero for broad values of $\Gamma$, i. e., FoG orbits are sensitive to initial conditions for large combinations between $\Gamma$ and $\mu$. Moreover, the interval $[0,1]$ is fully occupied when $\mu \rightarrow 4$, which we will discuss in the next section.

The use of the FoG map as a PRNG is significant in several ways: the large variety of parameters that exhibit chaotic behavior, the use of a really simple equation and outputs in the interval $[0,1]$ without the need for proper parameterization.

\section{Dynamical analysis FOG for \texorpdfstring{$\mu\rightarrow 4$}{Lg}}
\label{sec3}

In this section, we cover the case $\Gamma \rightarrow 4$ in more detail, showing the frequency distribution, the cobweb diagram, the time-evolution, the bifurcation diagram, and the Lyapunov exponent, in order to study the potential as PRNG and to obtain more information about the equation behavior.

\subsection{Time-evolution FoG for \texorpdfstring{$\mu\rightarrow 4$}{Lg}}

\begin{figure}[!hbtp]
\centering
\includegraphics[width=\columnwidth]{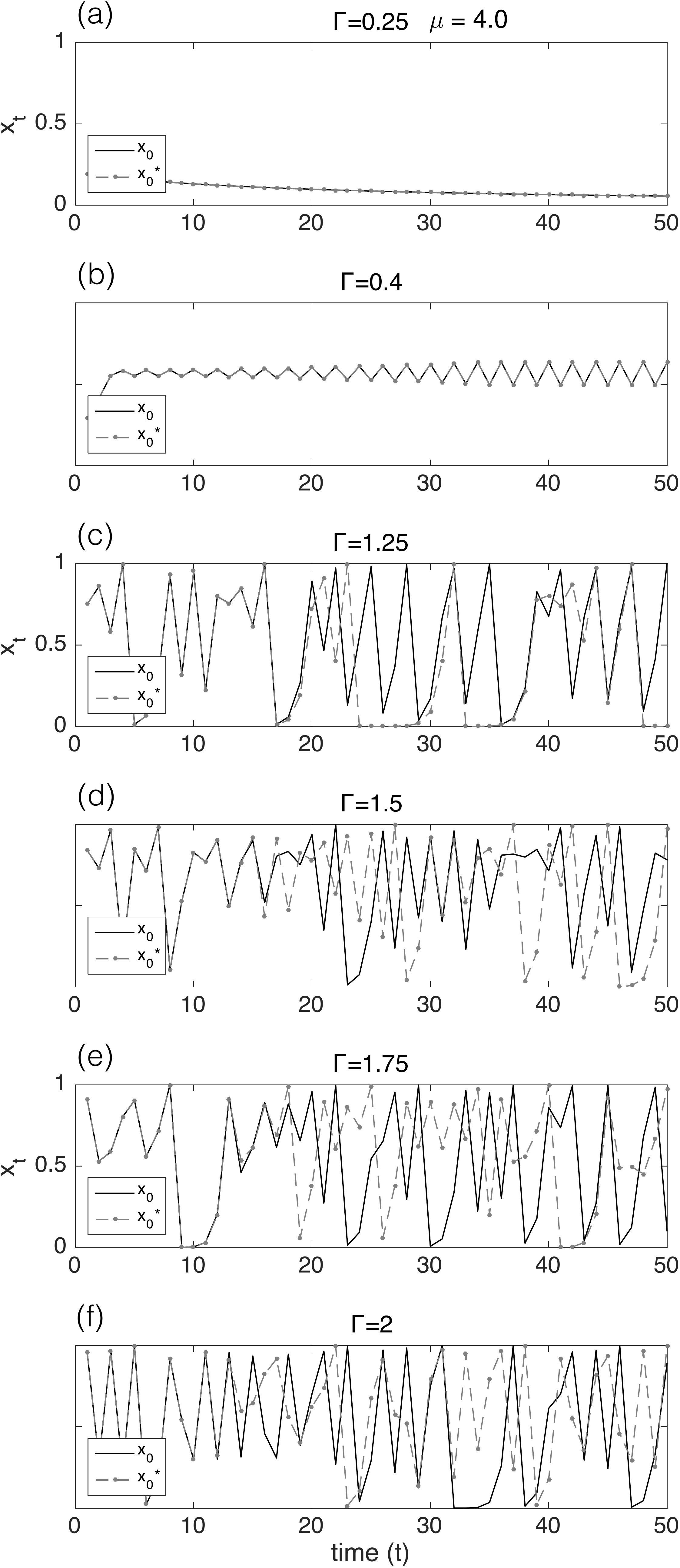}
\caption{Time-evolution of two orbits with close initial conditions $x_0=0.80000000$ (solid line) and $x_0^{\prime}= 0.80000001$ (dotted line) using parameter $\mu=4.0$ and (a) $\Gamma=0.25$, (b) $\Gamma = 0.40$, (c) $\Gamma = 1.25$, (d) $\Gamma = 1.50$, (e) $\Gamma = 1.75$ and (f) $\Gamma = 2.00$, over $t=50$ iterations.}
\label{fig:time_u4} 
\end{figure}

Fig.~\ref{fig:time_u4} shows the time-evolution of the composite map $FoG$ (Equation~\eqref{eq:fog}), where two orbits with close initial conditions evolved during $t=50$ iterations. Here, we used the specific parameters $\mu = 4.00$ and $\Gamma \in \{0.25, 0.40, 1.25, 1.5, 1.75, 2.00\}$. This plot shows three well-known Lyapunov stability behaviors, for example, the orbits are stable for $\Gamma = 0.25$ (Fig.~\ref{fig:time_u4}a). On the other hand, in (Fig.~\ref{fig:time_u4}b), the orbits for $\Gamma = 0.40$ show periodic behavior. Moreover, we can observe that two closely spaced orbits with initial conditions $x_0$ and $x_t^{\prime}$ lead to a chaotic regime when using parameter $\mu=4$ and $\Gamma>1$ (Figs.~\ref{fig:time_u4}c-f).

\subsection{Bifurcation Diagram FoG \texorpdfstring{$\mu\rightarrow 4$}{Lg}}

\begin{figure*}[!hbtp]
 \centering
 \includegraphics[width=\textwidth]{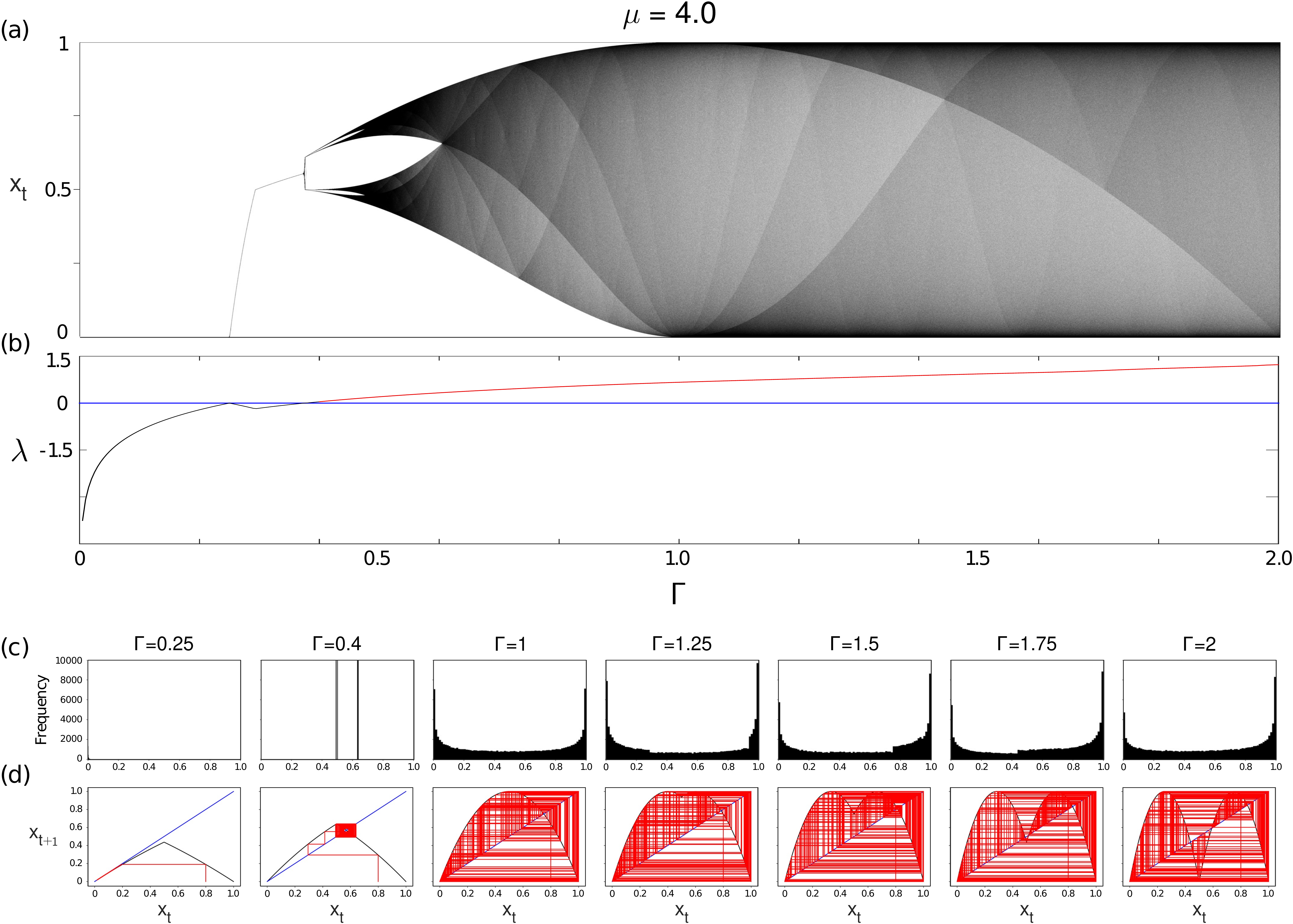}
 \caption{a) Bifurcation diagram of the FoG map with parameter $\mu=4.00$. The horizontal axis represents the parameter $\Gamma \in [0,2]$. The vertical axis $x_t \in [0,1]$ represents the long-term evolution for the same initial condition using the respective parameters $\mu$ and $\Gamma$. The system was iterated for $10^5$ iterations and the first 200 (transient time) were discarded. b) The corresponding Lyapunov exponent of (a) using Equation~\eqref{eq:normalLE}. The negative values ($\lambda <0$) are shown in black, while positive Lyapunov values ($\lambda\geq0$) are in red and the blue line indicates ($\lambda=0$). (c) Frequency distribution for various parameters $\Gamma=\{0.25, 0.4, 1.0, 1.25, 1.5, 1.75, 2.0\}$ and $\mu=4.0$. The horizontal axis shows the $x \in [0, 1]$ (500 bins) and the vertical axis shows the frequency of  $10^5$ values excluding the first $400$ transient values. d) The corresponding Cobweb plot for the same parameters in (c) shows the $x_t$ versus $x_{t+1}$ in red. The blue diagonal line represents the identity function, while the inverted parabolic curve in black shows the FoG function.}
 \label{fig:fog_full_u4}
\end{figure*}

For a value of $ \mu = 4 $, the orbits of the FoG map occupy the entire space $ [0,1] $ for values of $\Gamma$ larger than approximately $1$. Moreover, as Fig.~\ref{fig:fog_full_u4} shows, there is a high density of points in the regions where $x_t$ is close to $0$ and $1$. This feature is also present in the logistic map, indicating a preservation of this feature on the FoG map. Another observation about the density of orbits is the presence of several density ``contrasting curves", which are border regions between an area with low orbit density and an area with high orbit density. Note the contrasting curve running from the region of $x_t = 1$ to $x_t = 0$ separates two regions of higher orbit density at the top and less dense at the bottom of the bifurcation diagram, which is also observed in the frequency distribution histograms. Either we have $\Gamma \rightarrow 2$ darker (denser) regions for orbits near $1$ and for orbits near to $0$ it is less dense compared to values below than this value of $\Gamma$, while orbits between $ 0 $ and $ 1 $ are more homogeneous with respect to lower values of this $ \Gamma $, where the constraints curves are practically absent. Another feature seen in the bifurcation diagram is the presence of a division into four orbits immediately after the bifurcation into two orbits.

\subsection{Lyapunov stability FoG \texorpdfstring{$\mu\rightarrow 4$}{Lg}}

The Lyapunov exponent for the FoG map with $\mu = 4$ increases as $\Gamma$ increases, except for a region near to $\Gamma = 0.25$, where the Lyapunov exponent decays, but soon after it continues to grow, as we can see in Fig.~\ref{fig:fog_full_u4} (b). Another visible feature is the presence of positive Lyapunov exponents for most values of $\Gamma$, with the values of the Lyapunov exponent being positive at about $\Gamma>0.4$, reaching values close to $1.5$.

There are few values of $\Gamma$ where the Lyapunov exponent has values equal to zero. In Fig.~\ref{fig:fog_full_u4} (b) it can be seen that the Lyapunov exponent has values equal to zero for two values of $\Gamma$. That is, there are only a few cases that in which the orbits behave periodically. One of these cases was observed in Fig.~\ref{fig:time_u4} (f).

Moreover, in the region where the orbits occupy the whole range $[0,1]$, the Lyapunov exponent grows monotonically as the value of $\Gamma$ increases.

\subsection{Frequency distributions FoG \texorpdfstring{$\mu\rightarrow 4$}{Lg}}
\label{sec:freqdistrib}

Fig.~\ref{fig:fog_full_u4}c shows the frequency distributions for $\mu=4.0$ and various $\Gamma \in \{0.25, 0.40, 1.00, 1.25, 1.50, 1.75, 2.00\}$ with random initial conditions. The parameters $\Gamma = 0.25$ and $\Gamma = 0.40$ cover the trivial zone. And for $\Gamma \in \{1.00, 1.25, 1.50, 1.75, 2.00\}$, the orbits are more frequent at the ends, so the histogram havs the shape of a ``U''.

Also, there is a fraction in the ``U'' format for distributions where $ \Gamma \in \{1.25, 1.5, 1.75 \} $. For the distribution of $ \Gamma \in \{1.25, 1.5\} $, this shape break is near the value $1$ in space-phase and, for $\Gamma = 1.5$, this contrast is more in the middle of the frequency distribution. This is a consequence of the contrasting curves observed in the bifurcation diagram.

\subsection{Cobweb plot FoG \texorpdfstring{$\mu\rightarrow 4$}{Lg}}

The cobweb diagram is another way to analyze the stability and behavior of orbits generated by chaotic maps. In the case of the FoG map, the cobweb diagram with $ \Gamma \in \{1.0, 1.25, 1.5, 1.75, 2.0\}$ of Fig.~\ref{fig:fog_full_u4}d shows chaotic behavior that mirrors what was observed in the other forms of analysis of Fig.~\ref{fig:fog_full_u4}.
  
\subsection{Comparison between F, G and FoG}

\begin{figure*}[!hbtp]
 \centering
 \includegraphics[width=\textwidth]{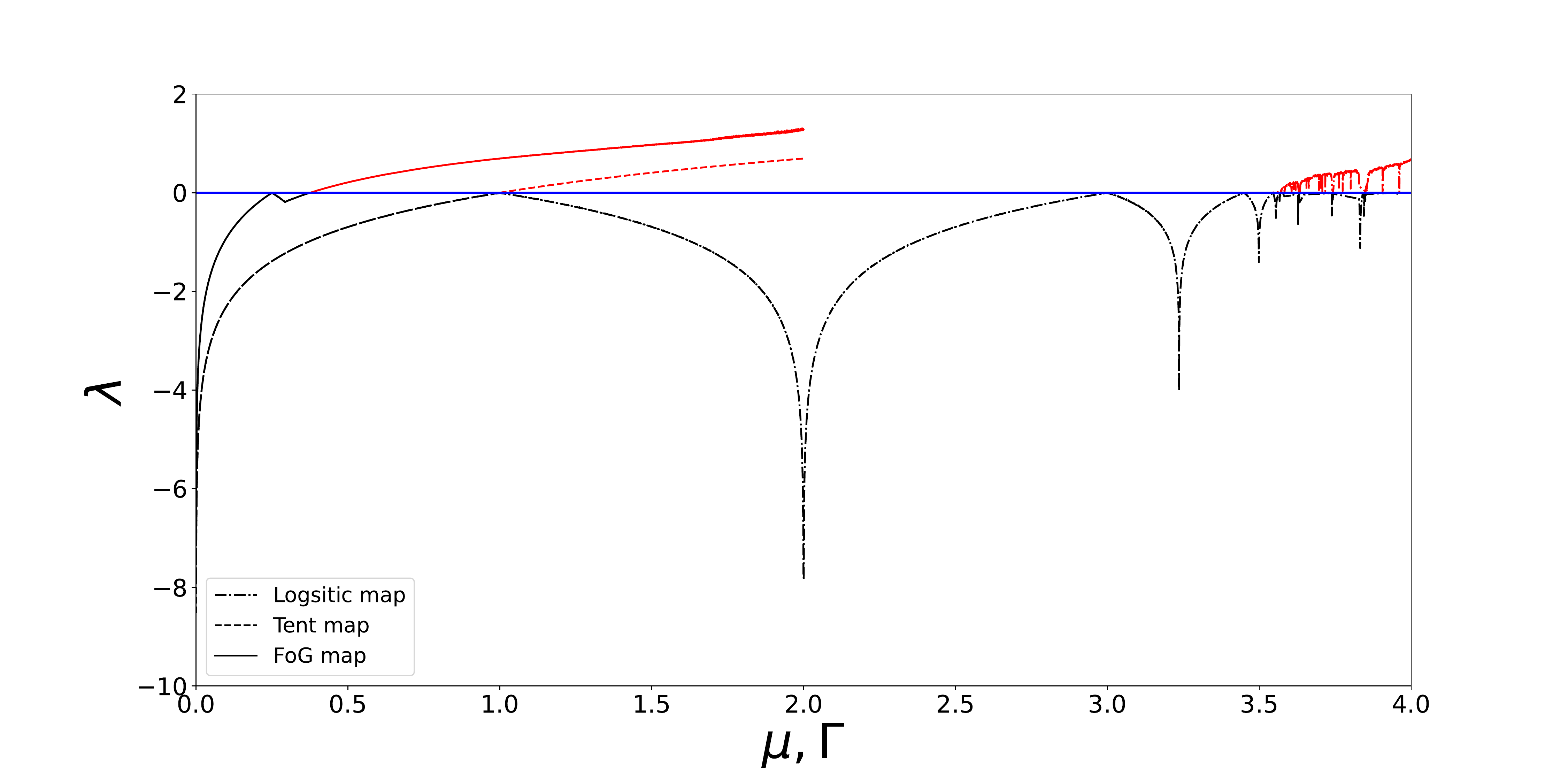}
 \caption{Comparison between the Lyapunov exponent of the FoG map, the logistic map and the tent map. }
 \label{fig:lyapunovs}
\end{figure*}

Fig.~\ref{fig:lyapunovs} compares the Lyapunov exponent of the tent map, the logistic map, and the composite map. It can be seen that the FoG map achieves higher values than the logistic map and the tent map. Nevertheless, the FoG has a higher Lyapunov exponent than the tent map. The Lyapunov exponent of the FoG map reaches values close to $1.5$ in $\Gamma = 2$ while the Lyapunov exponent of the logistic map and tent map reach lower values, approximately $ 0.69 $. 
Another feature of the FoG map is the ``U'' shape of the histogram of the frequency distribution, which is also found in the logistic map, indicating a conservation of this property.

Moreover, with $\mu = 4$ in terms of the important values of the orbits along the space-phase, the FoG map  generates orbits that fill the entire range $[0,1]$ for a wide range of values $\Gamma$, as can be seen in the bifurcation diagram. However, for the logistic and tent maps, their orbits fill the space-phase $[0,1]$ only for control parameters close to or equal to their maximum value: $\Gamma = 2$ for the tent map, and $\mu = 4$ for the logistic map.

\section{Deep-zoom analysis of \texorpdfstring{$k-FoG$}{Lg} for \texorpdfstring{$\mu=4.00$}{Lg}}
\label{sec4}
%This randomization method has been applied to the logistic map and tent map and were extended as the $k$-logistic map and $k$-tent map PRNGs.
The deep-zoom~\cite{Machicao17} is a randomization method that relies on the dismissed digits after the decimal separator. This randomization method takes, as input, a given orbit $\mathcal{O}$ from a chaotic system and removes the first $k$-digits of the fractional part of each point of the original orbit, yielding a new orbit $\mathcal{O}^k$. In this earlier work~\cite{Machicao17}, the relationship between the integer parameter $k$ and the quality of the pseudo-random sequences was demonstrated. This showed a rapid transition of the randomness from ``weak'' to ``strong'' as $k_{\geq0}$ increases. Thus, in the present work, we also intend to apply the deep-zoom approach to improve the randomness of the composite map.

%the deep-zoom can be applied to unimodal maps of the form 
%preserves phases-space as unimodal
Formally, the FoG map (Equation~\eqref{eq:fog}) is of the form $FoG(\mu,\Gamma):[0,1]\rightarrow [0,1]$ such that $\mathcal{O}(\mu, \Gamma,x_0)= \{x_0, x_1,\ldots, x_t\}$, or simply $\mathcal{O}^{k=0}$, is an orbit belonging to the phase-space of the FoG map with initial condition $x_0$. The deep-zoom approach applied to the FoG, hereafter referred to as the $k$-FoG map, yields a new orbit of the form $\mathcal{O}^{k}(\mu, \Gamma,x_0)= \{x_0^k, x_1^k,\ldots, x_t^k\}$ derived from the original orbit. The new value $x_t^k$ is formed by keeping the $k$-th decimals places to the right of the decimal separator corresponding to the underlying point $x_t \in \mathcal{O}^{k=0}$.

The $k$-FoG map is defined as a function given by Eq.~\eqref{eq:kfog} of the form $\phi_{k}(x_t):[0,1]\rightarrow ]0,1[$. The deep-zoom preserves the domain of the phase-space is $]0,1[$. This function generates orbits of the form $\mathcal{O}^{k}(\mu,\Gamma,x_0)= \{x_0^k, x_1^k,\ldots, x_t^k\}$, i.e., the initial orbit derived from the original orbit $\mathcal{O}(\mu,\Gamma,x_0)$. The new values $x_t^k$ are formed by keeping $k$-th decimals places to the right of the decimal separator corresponding to the underlying point $x_t \in \mathcal{O}^{k=0}$, a transformation that can be represented as follows:

\begin{equation}
	x_t^k= \phi_{k}(x_t) = x_t 10^{k} - \lfloor x_t 10^k \rfloor\,,
	\label{eq:kfog}
\end{equation}
where $\lfloor$ $\rfloor$ stands for the floor function, and $x_0 \in ]0,1[$. Machicao \& Bruno \cite{Machicao17} recommended to use $\mu \rightarrow 4$, since the closer $\mu$ is to 4, the less dense are the periodic windows, and therefore, a parameter taken by chance has a larger positive probability of generating a chaotic orbit.

\subsection{Bifurcation diagram for k-FoG}

\begin{figure}[!htbp]
 \centering
\includegraphics[width=\columnwidth]{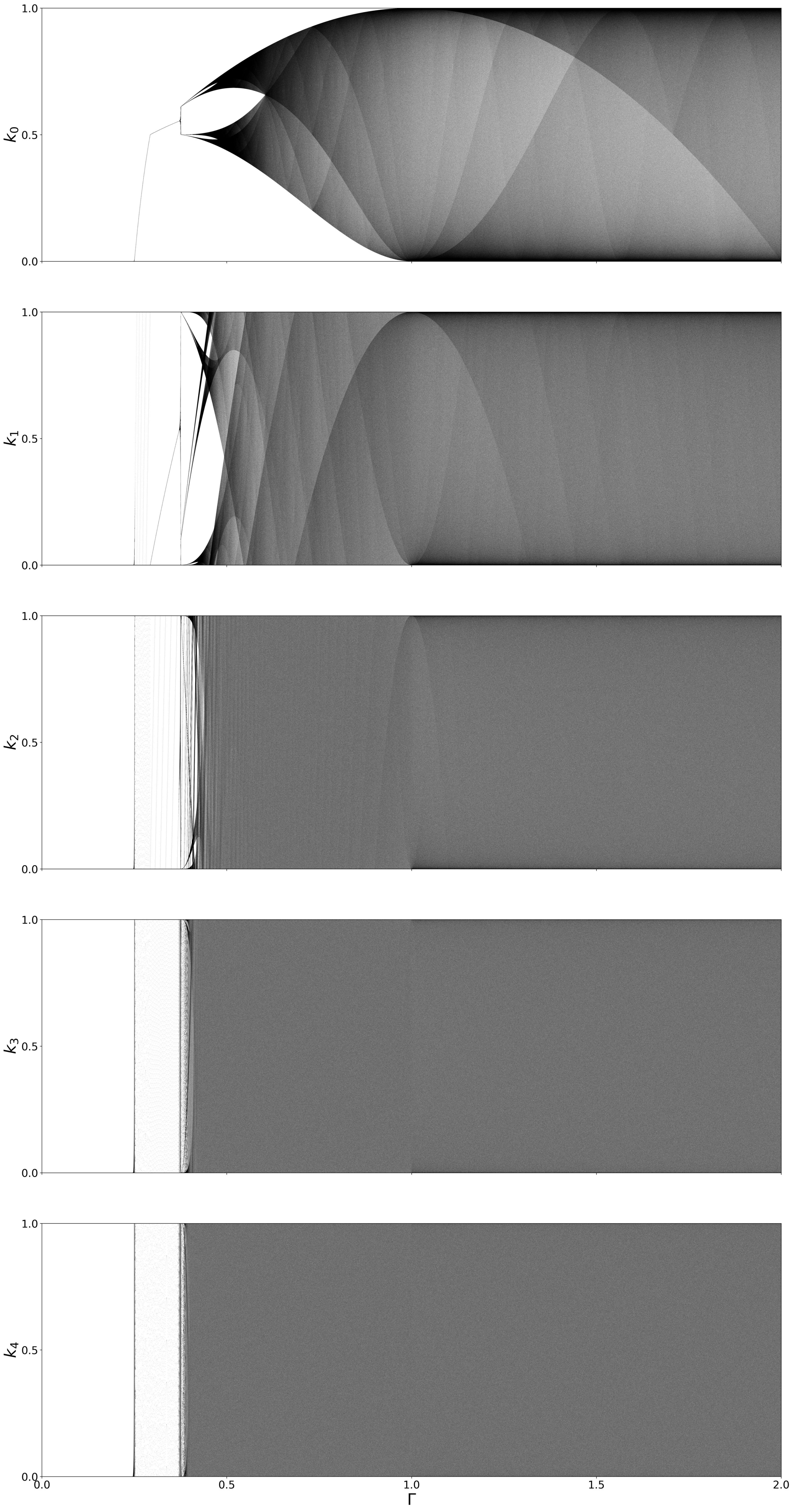}
\caption{Bifurcation diagrams of deep-zoom (Equation~\eqref{eq:kfog}) applied to the FoG map. From top to bottom, the $k_0$, $k_1$, $k_2$, $k_3$ and $k_4$-FoG map. The horizontal axis depicts $\Gamma \in [0,2]$ (steps of $0.003$). The vertical axis shows the possible long-term values of the corresponding $k$-logistic map, starting from the same initial condition, calculated over $10^5$ iterations and ignoring first $200$-th steps.}
\label{fig:bifkfog}
\end{figure}

The FoG map shows significant changes in the bifurcation diagram when the deep-zoom method is applied, such as: the distribution of orbits in space $[0,1]$ and the loss of ``contrasting curves''. Fig.~\ref{fig:bifkfog} shows a clear change in the number of orbits, where the interval $ [0,1] $ is more occupied as $k$ increases. This change in orbits leads to a loss of features in the frequency distribution, where the ``contrasting curves" become less visible and the dark regions near to $x_t = 1$ and $x_t = 0$ disappear as $k$ increases.

Since the deep-zoom method aims to make the orbits appear more random, as seen in the bifurcation diagram, the $k$-FoG map loses its features recognisable by the bifurcation diagram. The lack of information suggest that the deep-zoom randomization procedure is very efficient, considering how difficult it is to detect these features in the $k$-maps with larger values of $k$.

One possibility offered by the $k$-maps FoG is the exploration of more parameters of $ \Gamma $, such as orbits filling the space $[0,1]$ starting from $ \Gamma $ close to $ 0.5 $, while this phenomenon occurs on the FoG map without deep-zoom only for values of $\Gamma$ larger than approximately $1$. Thus, there are more possible combinations between $\Gamma$ and $\mu$ with the chaotic orbits, a property that can be explored, for example, in cryptography and PRNG.

\section{PRNG based on chaotic FOG with $\mu$=4 and $\Gamma$=2}
\label{sec5}

The $k$-FoG with given initial condition $x_0$ (seed) and real-valued hyper-parameter $\mu$, and positive integers $k$ and $d$, compute $\{x_n\}$ according to Eq.~\eqref{eq:Logistic}, can be formalized as follows \cite{Machicao17}:

\begin{equation}\label{eq:kchaoticmap}
x_n^k = x_n \cdot 10^{k} - \lfloor x_n \cdot 10^{k} \rfloor\,,
\end{equation}
where $\lfloor\cdot\rfloor$ is the floor operation. Observing that, the $x_n^k$ values remain in the unit interval. After that, each $x_n^k$ value is discretized by using the parameter $d$ in the following manner:

\begin{equation}\label{eq:discretized}
R_n= \lfloor x_n^k \cdot 10^d\rfloor \,.
\end{equation}

The resulting series $\{R_n\}_{n>0}$ is a sequence of pseudo-random integers in the range $0$ to $10^d-1$ generated by the $k$-logistic map PRNG. In our tests, we used \mbox{$d=1$}. The quality of this generated series is the subject of this present study.

In the context of chaos-based PRNGs, it is common to associate positive Lyapunov exponents $\lambda(x_0)>0$ with pseudo-random properties. The same relation was even proved in a paper by Lee et al.\cite{Lyapunov-GoodRandom}. Based on this argument, accessing the measure of sensitivity to initial conditions is of great interest to generate pseudo-random numbers. Indeed, using the Equation~\eqref{eq:kchaoticmap}, it has been shown that several chaotic regions of the logistic map have a positive Lyapunov exponent, as can be seen in Fig.~\ref{fig:bifkfog}. However, the positive Lyapunov exponent is not the only criterion required to use a chaotic system as a source of pseudo-randomness, as other criteria such as the frequency distribution and correlation analysis $x_t$, $x_{t+1}$ and $x_{t+2}$ should also be taken into accoutn, which are discussed more in detail in the following sections. 

\subsection{Statistical tests DIEHARD and TestU01 suits}

%DIEHARD
For the evaluation of the proposed PRNG, 100 files were generated for different $k$ paramter values, each containing 11.2 MB which corresponds to $28 \times 10^5$ numbers generated for each $k= \{0,1, \ldots, 9\}$. Note that the file size used is due to limitations of DIEHARD~\cite{DIEHARD} itself. Each sample was generated from a random seed, and each $x_{t}^{k}$ value of each orbit was transformed into integers using a $32$ bit mask (discretization) by converting it to the Equation~\eqref{eq:discretized}.

In this way, the test battery DIEHARD~\cite{DIEHARD} was applied to the generated dataset, assuming the null hypothesis ``$H_0$: is a random sequence '' if the tests yields results between $0.0001 <\text{p-value} <0.9999 $ \cite{DIEHARD}.

A more rigorous interpretation of this earlier technique was proposed in \cite{Diehard3regions}, where the $p$-values distributed in the range $[0,1)$ were divided into three types of areas defined by the following limits: 
\begin{itemize}
\item Safe-Area:    $0.25 < p$-value $<0.75$. 
\item Doubt-Area:   $0.1  < p$-value $\leq 0.25$ or $0.75 \leq p$-value $<  0.9$.
\item Failure-Area: $0 \leq p$-value $\leq 0.1 $ or $0.9  \leq p-value \leq 1$.
\end{itemize}

The results of the DIEHARD experiment are listed in Table~\ref{tab:pvalues}. Each column indicates the quality of randomness in terms of the number of files whose $p$-values fall within the ``Safe'' and ``Doubt'' ranges and sum at most $100$.

% TestU01
One of the most complete battery of randomness tests is TestU01~\cite{l2007testu01}. This suite contains three types of batteries: SmallCrush, Crush and BigCrush, in which BigCrush was selected. This test battery includes a total of $106$ tests that require a resolution of at least $30$ bits and uses approximately $2^{38}$ numbers of the PRNG. The result of the test battery provides $p$-values that pass if the $p$-values are in the range of $[0.001, 0.999]$ and fail if they are in this range.

Furthermore, the generated chaotic map orbits for TestU01 were calculated with 512-bit floating-point and converted to IEEE double precision floating-point arithmetic \cite{IEEE}. In addition, the following control parameters were chosen: $\mu = 3.99999999$ and $\Gamma = 1.99999999$; and the k values analyzed are in the range 0 to 11.

\subsection{Comparison of PRNGs}
\label{sec:comparison}
 
\begin{figure}[!hbtp]
 \centering
 \includegraphics[width=\columnwidth]{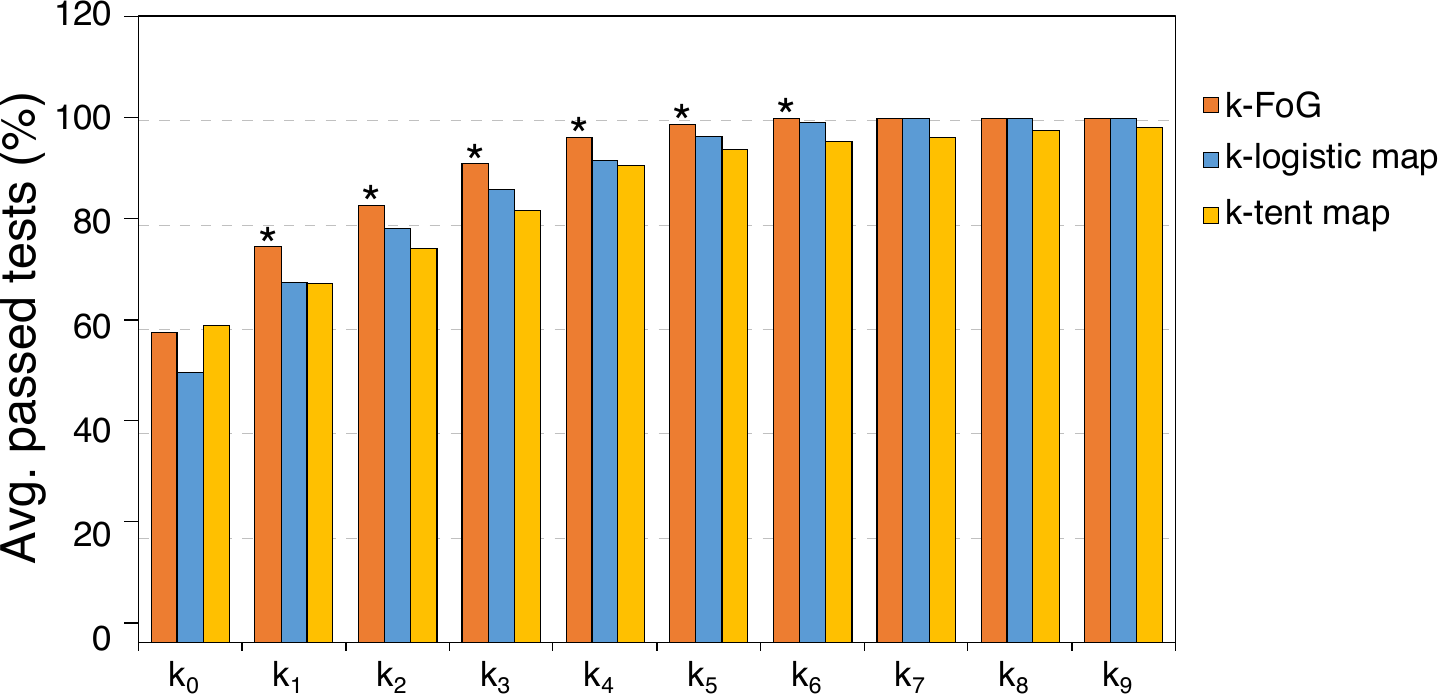}
 \caption{Comparison DIEHARD for $k$-FoG, $k$-logistic map, and $k$-tent map PRNG. Number of files that passed the DIEHARD test of $k$-logistic map, $k$-tent map for $k_0$ to $k_{9}$ and $k$-FoG PRNGs. The vertical axis shows the average number while the horizontal axis gives the different values of the $k$-digits. The curves represent the average number of tests passed. The asterisk represents the PRNG that performed best in each case when using the Wilcoxon T-test.}
 \label{fig:comparison}
\end{figure}

To compare the performance of the pseudo-randomness properties of the $k$-logistic map and the $k$-tent map with the $k$-FoG. The average number of files that passed the DIEHARD test is summarized in Fig.~\ref{fig:comparison}. This plot shows comparable results obtained for these PRNGs.

To verify a comparison between the PRNGs, we used the Wilcoxon T-test for comparing paired data samples. This is a non-parametric statistical hypothesis test that can be used to determine whether two data samples have the same distribution or a different distribution.

\begin{figure}[!hbtp]
 \centering
 \includegraphics[width=\columnwidth]{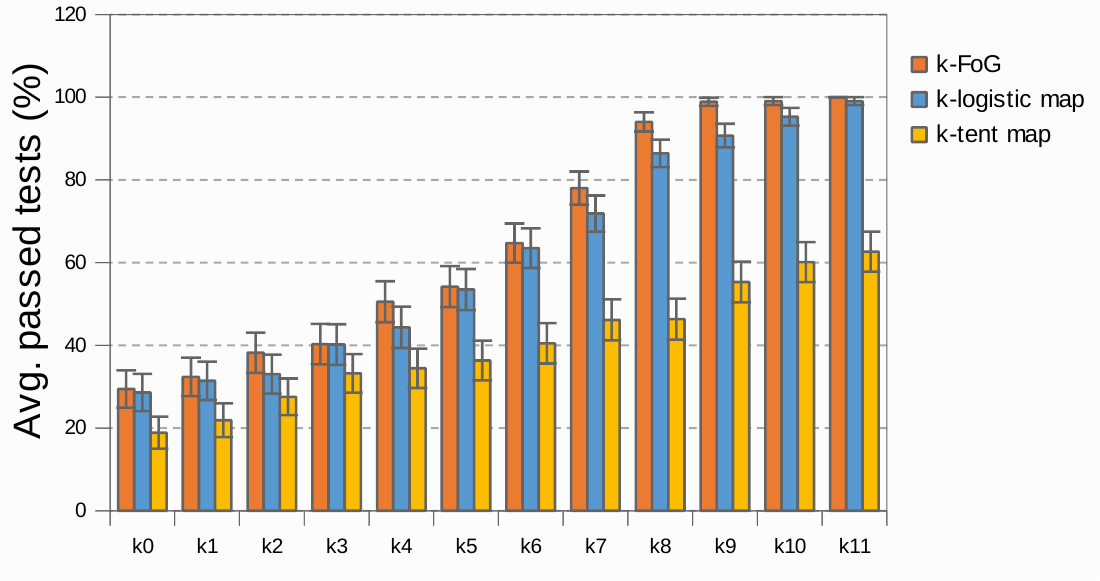}
 \caption{Comparison TestU01 BigCrush for $k$-FoG, $k$-logistic map, and $k$-tent map PRNG. The vertical axis shows the average number, while the horizontal axis gives the different values of the $k$-digits. The curves represent the average number of tests passed.}
 \label{fig:comparison-testu01}
\end{figure}

The results of the TestU01 BigCrush battery for the $k$-FoG map, $k$-logistic map and $k$-tent maps is shown in Fig.~\ref{fig:comparison-testu01}. Some relevant observations to the figure are: the $k$-FoG passes in more tests than the $k$-logistic map and $k$-tent map for all analyzed $k$ values.

% Table \ref{tab:pvalues}
 
 \begin{table}
     \centering
     \includegraphics[width=\columnwidth]{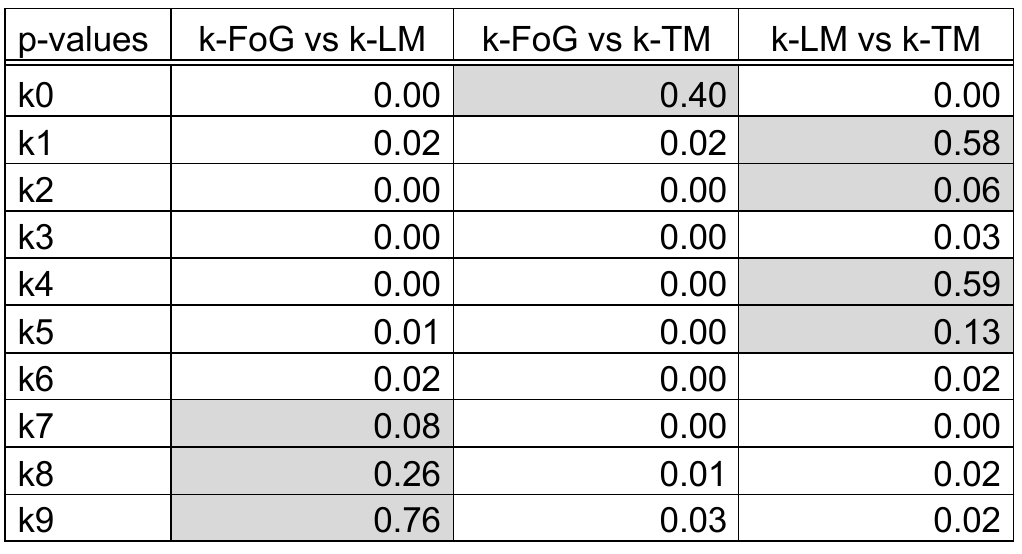}
     \caption{P-values for the comparison of pairs obtained by the Wilcoxon T-test.}
     \label{tab:pvalues}
 \end{table}
\section{Discussions}
\label{sec6}

The FoG map showed a large variety of orbits and consequently, a large variety of bifurcation diagrams and Lyapunov exponent values, as can be seen in Fig.~\ref{fig:fog_full_u4}. In it, for some values of $ \mu $, the FoG map shows no chaotic behavior ($ \mu \in [0,1.79] $ and $ \mu \in [2.93, 3.25] $), while for other values of $ \mu $ the chaotic behavior is visible ($ \mu \in [1.79, 2.93] $ and $ \mu \in (3.25, 4.0] $). For $ \mu \rightarrow 4 $, the FoG map showed interesting features such as: increased Lyapunov exponent and the presence of orbits throughout the entire $[0,1]$ range. These features are relevant for the use as PRNG, since the Lyapunov exponent of the FoG map reaches higher values than the logistic map and tent map, we have that the orbits of the FoG map digest exponentially faster than the isolated maps. And filling the interval $[0,1]$ by the orbits is one of the basic requirements of a PRNG, where the FoG map has this property for large values of $\Gamma$, approximation for $\Gamma > 1$.

The $k$-FoG maps created using the deep-zoom method changed the orbits in a way that appeared increasingly random as $k$ increased. This seemingly random appearance was reflected in the bifurcation diagram, where its features in the densities of the orbits, with the ``contrasting curves" perpendicular to the increase of $k$, and the orbits of the $k$-FoG map occupy the entire $[0,1]$ of $\Gamma$ close to $0.5$ as $k$ increases.

To quantify the random occurrence of the orbits, the $k$-FoG maps passed the statistical test battery DIEHARD, TestU01 and compared the result with the results of the $k$-logistic maps and $k$-tent maps. The comparison showed that the $k$-FoG map obtained better test results than the $k$-logistic map and $k$-tent map for most $k$ values. In this way, the $k$-FoG map proved to be particularly suitable for PRNG and cryptography applications.  

\section{Conclusion}
\label{sec7}

The obtained results showed that the composition $(f\circ g)(x_t)$ has the potential to improve properties of PRNG, especially for control parameters $ \mu $ close to $4$. For this condition, the FoG map orbits achieved higher Lyapunov values higher than the Lyapunov exponents of the logistic map and tent map orbits. And, the bifurcation diagram of the FoG map for such conditions shows a wide range of values for $\Gamma$ in which the orbits occupy the entire space-phase $[0, 1]$. Applying the deep zoom method, the orbits occupy the range $[0, 1]$ for most of the $\Gamma$ values as $k$ increases. Moreover, as $k$ increases, the FoG map apparently generates more random orbits, as illustrated by the bifurcation diagrams and confirmed by the DIEHARD and TestU01 battery of statistical tests.

\section*{Acknowledgments}
J. P. V. acknowledges support from S\~ao Paulo Research Foundation FAPESP (grant \#21/07377-9).
J. M. is grateful for the support from FAPESP (grant  2020/03514-9). 
O. M. B. acknowledges support from CNPq (grant \#307897/2018-4) and FAPESP (Grant \#18/22214-6). 

% \section*{References}

%----------------------------------------------------

\bibliographystyle{elsarticle-num}
%\bibliographystyle{elsarticle-num-names-alphsort}
%\bibliography{biblio}

\end{document}